\documentstyle[12pt,epsfig]{article}

\newcommand{\beq}{\begin{equation} }
\newcommand{\eeq} {\end{equation} }

\begin{document}

\begin{flushright}
{\tt hep-ph/0212339}\\
OSU-HEP-02-18\\
December 2002\\
\end{flushright}

\vspace*{2.5cm}
\begin{center}
{\Large {\bf Stabilizing the Axion by Discrete Gauge Symmetries
 } }

\vspace*{2.5cm}

 {\large {\bf K. S. Babu,\footnote{E-mail address: babu@okstate.edu}
 Ilia Gogoladze\footnote{On a leave of absence from:
Andronikashvili Institute of Physics, GAS, 380077 Tbilisi,
Georgia.  \\ E-mail address: ilia@hep.phy.okstate.edu}} \bf and
Kai Wang\footnote{E-mail address: wk@okstate.edu}

 \vspace*{1cm}

{\small \it Department of Physics, Oklahoma State University\\
Stillwater, OK~74078, USA }}
\end{center}

\begin{abstract}
The axion solution to the strong CP problem makes use of a global
Peccei--Quinn $U(1)$ symmetry which is susceptible to violations from
quantum gravitational effects.  We show how discrete gauge
symmetries can protect the axion from such violations.
PQ symmetry emerges as an approximate global symmetry from
discrete gauge symmetries. Simple models based on
$Z_N$ symmetries with $N=11,12$, etc are presented realizing the
DFSZ axion and the KSVZ axion.  The discrete gauge anomalies are
cancelled by a discrete version of the Green--Schwarz mechanism.
In the supersymmetric extension our models provide a natural link
between the SUSY breaking scale, the axion scale, and the
SUSY--preserving $\mu$ term.
\end{abstract}

\newpage

\section{Introduction}

One of the challenges facing the Standard Model is an understanding of the
strong CP problem \cite{kim}. The Lagrangian of quantum chromo-dynamics (QCD)
contains a term
\beq
\label{qcd}L_{QCD}\supset\theta\frac{g^2_s}{32\pi^2}~G_{\mu \nu}\tilde{G}^{\mu\nu},
\eeq
where  $G^{\mu \nu}$
is the gluon  field strength tensor, $g_s$ the QCD gauge coupling,
and $\theta$ a dimensionless  parameter. This
term violates both P and CP invariance.  In the full QCD Lagrangian
involving massive quarks,  $\theta$ is not a physical observable,
but $\bar{\theta}$ defined as
\beq \label{qcd1} \bar{\theta}=\theta +
\mathrm{arg}(\mathrm{det}M_U~ \mathrm{det}M_D), \eeq
is.  Here $M_U$ and $M_D$ are the up--quark and the down--quark  mass
matrices. The observed CP violation in weak interactions indicates
that the complex phase in the Cabibbo--Kobayashi--Maskawa
matrix is of order unity.  Without additional symmetries one would
expect that the quark mass matrices, and hence their determinants,
also contain large complex
phases. Therefore one would naturally expect from Eq.
(\ref{qcd1}) that $\bar{\theta}$ is order one. However, this is
in gross violation of the experimental limits arising from the
neutron electric dipole moment (EDM).  Eq. (\ref{qcd1}) leads to a
neutron EDM of order $d_n\simeq
5\times10^{-16}~\bar{\theta}$ e cm. Current experimental limit on $d_n$ is
$d_n<10^{-25}$ e cm \cite{pdg},
which puts a constraint  $\bar{\theta}<
10^{-10}$.  Why $\bar{\theta}$ is so small, when its natural values if
of order one, is the strong CP problem.

An elegant and popular solution to the strong CP problem is provided by
the Peccei--Quinn mechanims \cite{PQ}.  It is based on
on a global axial symmetry, $U(1)_{PQ}$,
which  is spontaneously broken.  In this mechanism
$\overline{\theta}$ is promoted to a  dynamical field,
$\overline{\theta}=a/f_a$, with an effective potential for this field induced by
non--perturbative QCD effects.
Here $a$ stands for the pseudo--Goldstone mode
of the spontaneously broken $U(1)$ symmetry, the axion \cite{WW}, and $f_a$ the
axion decay constant (equal to the VEV of the scalar
that breaks the PQ symmetry). Minimizing the axion potential would
fix the vacuum expectation value (VEV) of $a$, or equivalently
$\bar{\theta}$ to zero, thus
providing a natural solution to the strong CP problem.

The axion decay constant $f_a$ is model--dependent.  In the original
Weinberg--Wilczek axion \cite{WW}, the PQ symmetry is broken spontaneously at
the electroweak scale.  This model contains two Higgs doublets
$\Phi_1$ and $\Phi_2$ with VEVs $v_1$ and $v_2$.
Without the QCD instanton effects, spontaneous breaking of the global
$U(1)$ symmetry would imply that the axion is massless.  Including
non--perturbative QCD effects which break the global $U(1)$ explicitly
the axion would acquire a mass given by
\cite{WW}
 \beq
 \label{mm1}
m_a\sim \frac{f_{\pi}m_{\pi}}{v}\simeq 100~ \mathrm{keV}
 \eeq
where $m_{\pi}$ and $f_{\pi}$ are the pion mass and  decay constant,
and $v=\sqrt{v_1^2+v_2^2} \sim 246$ GeV.  The axion decay constant
is $f_a \sim v$ in this model. Laboratory data, especially from the
non--observation of the rare decay $K \rightarrow \pi + a$,
excludes this simple weak scale realization of the axion \cite{kim,pdg}.

The laboratory bound on the axion coupling can be evaded by introducing
a singlet scalar $S$ into the theory carrying a non--zero PQ charge.
When this scalar acquires a VEV, the global $U(1)$ symmetry will break
spontaneously.  The axion decay constant in this case is $f_a \sim \left
\langle S \right\rangle$, which can be much larger than the weak scale.
With $f_a \gg v$, the
couplings of the axion with the Standard Model fields will be highly
suppressed,
by a factor $v/f_a$, and the laboratory limits are evaded.  This class
of models are the ``invisible axion models".  If the spectrum of the
theory contains the scalars $S$ and two doublets $\Phi_1$ and $\Phi_2$,
we have Dine--Fischler--Srednicki--Zhitnitskii (DFSZ) axion \cite{DFSZ}.  If the
scalar spectrum contains a single Higgs doublet and a singlet $S$
but the fermionic spectrum  is enlarged to contain a pair of quark--antiquark
fields, we have the Kim--Shifman--Vainshtein--Zakharov (KSVZ) axion \cite{KSVZ}.
In the DFSZ axion model the SM fermions carry PQ charges and the axion potential
is induced by the QCD chiral anomaly associated with the SM fermions.
In the KSVZ axion model only
the exotic vector--like quarks carry PQ charges and their QCD anomaly
induces the axion potential.

The PQ symmetry breaking scale $f_a$ is constrained by a combination of
laboratory, astrophysical, and cosmological limits to be in the
range $10^{10}$ GeV $\leq f_a \leq 10^{12}$ GeV.  If $f_a$ is much above
the weak scale, axions produced in the interior of stars can escape freely,
draining the star of its energy too rapidly.  The limit $f_a \leq 10^{10}$ GeV
arises from the stability of stars and supernovae \cite{kim}.  If $f_a$
exceeds about $10^{12}$ GeV, the energy density in the coherent oscillations
of the axions will over-close the universe, contradicting cosmological limits.
The mass of the axion in these invisible models is extremely small, $m_a
\sim f_\pi m_\pi/f_A \sim 10^{-4}$ eV.  Axion with a mass in this range
can constitute a good cold dark matter candidate \cite{kim}.

Quantum gravitational effects can potentially violate the global PQ symmetry
which is needed for the axion solution. These effects, associated with black holes, worm holes, etc
are believed to violate all global symmetries, while they respect all
gauge symmetries \cite{hawking}.  When a gauge symmetry is spontaneously broken, often
a discrete remnant survives \cite{kw}.  Such discrete gauge symmetries are left intact
by quantum gravity.  Any such quantum gravity violations of the PQ symmetry
should be extremely small for axion  to be a viable solution to the
strong CP problem.  For example, in the invisible axion models, the gauge
symmetries would allow for a term in the scalar potential of the form
\beq \label{ax2}  V \supset \frac{S^n}{M_{\rm Pl}^{n-4}},
 \eeq
where $S$ is the singlet field carrying PQ charge, $M_{\rm Pl}$ is
the Planck scale, and $n$ is a positive integer.
This term would induce a non--zero $\bar{\theta}$
given by
\beq \bar{\theta} \simeq \frac{f_a^{n}}{{M_{\rm Pl}}^{n-4}\Lambda_{QCD}^{4}}
\eeq
where $\Lambda_{QCD} \simeq 0.2$ GeV is the QCD scale.  Since the
neutron EDM sets the limit $\bar{\theta} \leq 10^{-10}$, using
$f_a \sim 10^{11}$ GeV one finds that $n \geq 10$ is necessary in Eq. (4)
\cite{barr}. This is indeed a severe constraint on axion models.

A possible way to avoid the problems associated with quantum gravity
is to identify Peccei--Quinn symmetry as an approximate global
symmetry associated with a gauge symmetry.  Attempts have been
made in the past with some success
to achieve this by extending the low energy (below the Planck
scale) particle content of the invisible axion models \cite{barr}.
In this paper we show how it is possible to use discrete gauge symmetries
to stabilize the axion solution without enlarging the low energy
particle content.  This is made possible by a discrete version of
the Green--Schwarz anomaly cancellation mechanism \cite{gs}. Superstring
theory when compactified to four dimensions generically contains
an anomalous $U(1)$ symmetry with its anomalies cancelled by a shift
in the pseudoscalar partner of the dilaton field (the string axion field).
This anomalous $U(1)$ is broken near the string scale when some scalar fileds
carrying $U(1)_A$ charges are shifted to cancel the Fayet--Iliopoulos term
associated with the $U(1)_A$ (so that supersymmetry is left unbroken near
the string scale) \cite{dsw}.  Often a discrete $Z_N$ subgroup of the $U(1)_A$
symmetry is left unbroken to low energies.  This $Z_N$ symmetry, being
of a gauge origin, is protected against quantum gravitational violations.

From an effective theory point of view the conditions for Green--Schwarz
anomaly cancellations are
 \beq \label{gs}
\frac{A_{i}}{k_{i}}=\frac{A_{gravity}}{12}=\delta_{GS}, \eeq
where $A_i$ are the anomaly coefficients associated with $G_i^2 \times
U(1)_A$ with $G_i$ being one of the factors of the surviving gauge
symmetry. $\delta_{GS}$ is a constant that is not specified by
the low energy theory alone.  $k_i$ are the levels of
the  Kac--Moody algebra.  For non--Abelian groups $k_i$ are (positive)
integers, although for $U(1)$ factors $k_i$ need not be integers.
All other anomaly coefficients such as  $G_i G_j G_k$ and
$[U(1)_A]^2 \times G_i$ should vanish.
In the case of $SU(3)_C \times SU(2)_L \times U(1)_Y \times
U(1)_A$ being the surviving symmetry, the GS anomaly cancellation
conditions are
 \beq \label{gs11}
\frac{A_3}{k_3}=\frac{A_2}{k_2}=\delta_{GS}, \eeq
where $A_3$ and $A_2$ stand for the $[SU(3)_C]^2 \times U(1)_A$ and
$[SU(2)_L]^2 \times U(1)_A$ anomaly coefficients. There is an analogous
expression involving the hypercharge $Y$, however since $Y$ is not quantized,
the associated level $k_1$ is not an integer in general and this condition
is not very useful from an effective low energy theory point of view \cite{banks}.
The $[U(1)_A]^3$ anomaly can be cancelled by the GS mechanism, but this
condition also has an arbitrariness from the normalization
of $U(1)_A$.  The anomaly
in $[U(1)_A]^2\times  U(1)_Y$ also does not give any useful low energy constraint.

When the $U(1)_A$ symmetry breaks down to a $Z_N$ discrete symmetry,
the effective low energy theory would satisfy the following
discrete Green--Schwarz anomaly cancellation condition \cite{banks,ibanez1,bin,wk}:
\beq \frac{A_3+mN/2}{k_3}=\frac{A_2+m'N/2}{k_2} \eeq
where $m$ and $m'$ are integers.  Heavy particles which are chiral
under $U(1)_A$ can acquire masses of order the $U(1)_A$ breaking scale
through $Z_N$--invariant Yukawa couplings, the
contributions proportional to the integers $m$ and $m'$ account for
these heavy fermions.

In what follows, we shall impose  the conditions of Eq. (8), which
should be sufficient for the surviving $Z_N$ symmetry to be protected
from quantum gravitational effects.

\section{Stabilizing the  DFSZ Axion}

The non--supersymmetric DFSZ axion model \cite{DFSZ} introduces two Higgs doublets
$H_u$ and $H_d$ and a Standard Model singlet scalar $S$.  The Lagrangian
of the model relevant for the discussion of axion physics is
\beq \label{ds1}
{\cal L} = Qu^{c}H_u+Qd^{c}H_d+Le^{c}H_d+L\nu^{c}H_u+\nu^{c}\nu^{c}-
\lambda({H_u}H_{d}S^{2}+h. c).\eeq
Here we have used a standard notation that easily generalizes to
our supersymmetric extension as well.  $Q,L$ are the left--handed quark
and lepton doublets, $u^c, d^c, e^c, \nu^c$ are the quark and lepton
singlets.  The seesaw mechanism \cite{seesaw} for small neutrino
masses has been incorporated in Eq. (9).

Eq. (9) has three $U(1)$ symmetries, as can be inferred by solving
the six conditions imposed by Eq. (9) on nine parameters.
These three $U(1)$ symmetries can be identified as the SM hypercharge $U(1)_Y$,
baryon number $U(1)_B$ and a Peccei--Quinn
symmetry $U(1)_{PQ}$.
If we denote the charges of $(Q$, $u^c$,
$d^c)$ as $(q,~u,~d)$, the symmetries can be realized as $B=q-u-d$,
$PQ=-d$, $Y/2=q/6-2u/3+d/3$.  The $U(1)$ charges of the various
particles under these symmetries are listed in Table 1.

\begin{table}[ht]
 \begin{center}
  {\renewcommand{\arraystretch}{1.1}
  \begin{tabular}{| c | c  c  c  c  c  c  c  c  c  |}
     \hline

     \rule[5mm]{0mm}{0pt}
   & $Q$ & $u^{c}$ &
   $d^{c}$ & $L$ &
   $e^{c}$ & $\nu^{c}$ &
   $H_d$ & $H_u$ &
   $S$\\
   \hline
   \rule[5mm]{0mm}{0pt}
   Y/2 & 1/6 & $-2/3$ & 1/3 &
   $-1/2$ & 1 & 0 & $-1/2$ & 1/2 & 0\\
    \rule[5mm]{0mm}{0pt}
    $B$ & 1 & $-1$ & $-1$ & 0 & 0 & 0 & 0 & 0 & 0\\
    \rule[5mm]{0mm}{0pt}
   $PQ$ & 0 & 0 & $-1$ &
   0 & $-1$ & 0 & 1 & 0 & $-1/2$\\
   \hline
  \end{tabular}
  }
  \caption {\footnotesize  $Y/2$,
  $B$ and $PQ$ symmetries corresponding to hypercharge, baryon number and PQ
  charge respectively. The charges are assumed to be generation independent.}
  \label{MSSM}
 \end{center}
\end{table}

After $H_d$, $H_u$ and $S$ fields develop VEVs, the global PQ symmetry is
broken and the light spectrum contains a  Goldstone boson, the
axion. Non--perturbative QCD effects induce an
axion mass \cite{kim} given by
\beq
\label{ax1}
 m^{DSFZ}_a \simeq 0.6 \times 10^{-4}~ {\mathrm eV}~ \frac {10^{11}~
 \mathrm{GeV}}{f_a},
 \eeq
where $f_a \sim \left\langle S \right\rangle$ is the axion decay constant.

We now apply the Green--Schwarz mechanism for discrete anomaly cancellation
to stabilize the axion from quantum gravity corrections.  Even though the
model under discussion is non-supersymmetric, the GS mechanism for
anomaly cancellation should still be available, since SUSY breaking
in superstring theory need not occur at the weak scale in principle.
Since baryon number has no QCD anomaly, any of its subgroup will be
insufficient to solve the strong CP problem via the PQ mechanism.
On the other hand, the PQ symmetry does have a QCD anomaly, although with the
charges listed in Table 1 it has no $SU(2)_L$ anomaly.  Since hypercharge
$Y$ is anomaly free, we attempt to identify the anomalous $U(1)_A$ symmetry as
a linear combination of $PQ$ and $B$:
\beq
 \label{an1} U(1)_A=PQ + \gamma B.
 \eeq
According to the Eq. (\ref{an1}) and the charge assignment presented
in Table 1, we have for the anomaly coefficients for the $U(1)_A$,
\begin{eqnarray}
 \label{an2}
 A_3&\equiv & [SU(3)]^2\times U(1)_A=-\frac{3}{2}  \nonumber\\\
 A_2&\equiv & [SU(2)]^2\times U(1)_A= \frac{9}{2}~\gamma~.
\end{eqnarray}
If we identify $\gamma = -k_2/(3k_2)$, the anomalies in $U(1)_A$
will be cancelled by GS mechanism.  Thus we have
 \beq \label{an4}
U(1)_A=PQ-\frac{1}{3}\frac{k_2}{k_3}B. \eeq
The simplest possibility is $k_2=k_3 =1$, corresponding to the levels of
Kac--Moody algebra being one.
Normalizing the charge of the singlet field $S$ to be an integer,
Eq. (\ref{an4}) can be rewritten as
 \beq \label{an5}
U(1)_A=6 (PQ)-2(B). \eeq
The corresponding charge assignment is given in Table
\ref{charge5}. As discussed earlier, since hypercharge $Y$ is
anomaly free, one can add a constant multiple of $Y/2$ to the
$U(1)_A$ charges, and still realize GS anomaly cancellation mechanism.
The charges listed in Table 2 assumes the combination
$-\frac{5}{3}(6PQ-2B+\frac{4}{5}Y)$.  As can be seen from Table 2,
this choice of charges is compatible with $SU(5)$ grand unification.

Suppose that the $U(1)_A$ symmetry is broken near the string scale
by the VEV of a scalar field which has a $U(1)_A$ charge of $N$
in a normalization where all $U(1)_A$ charges have been made integers.
A $Z_N$  symmetry will then be left unbroken to low scales.  Two
examples of such $Z_N$ symmetries are displayed in Table 2 for $N= 11,12$.
Invariance under these $Z_N$ symmetries will not be spoiled by quantum
gravity, it is this property that we use to stabilize the axion.

Potentially dangerous terms that violate the $U(1)_{PQ}$ symmetry
of Eq. (9) are $S^n/M_{\rm Pl}^{n-3}$, $H_u H_d S^{* m}/M_{\rm Pl}^{m-2}$
etc, for positive integers $n,m$.  For the induced $\bar{\theta}$ to be
less than $10^{-10}$, the integers $n,m$ must obey $n \geq 10, m \geq 5$.
The choice of $N=11,12$ satisfy these constraints.  Note that  a $Z_{10}$
discrete symmetry would have allowed a term $S^2$, which would be
inconsistent with the limit on $\bar{\theta}$.  $Z_N$ symmetries with $N$ larger than 12
can also provide consistent solutions. Since by construction, the $U(1)_A$
symmetry in Table 1 is anomaly-free by GS mechanism,
any of its $Z_N$ subgroup is also anomaly--free by the discrete GS mechanism,
as can be checked directly.  In the $Z_{11}$ model, for example, we have
$A_3 = A_2 = 4$.  Consistent with the $Z_{11}$ invariance, terms that
violate the  $U(1)_{PQ}$ symmetry and give rise to an axion mass are
$S^{11}/M_{\rm Pl}^7,~H_uH_dS^{*9}/M_{\rm Pl}^7$ etc, all of which are quite harmless.
We conclude that the DFSZ axion
can be stabilized against potentially dangerous non--renormalizable
terms arising from quantum gravitational effects in a simple way.

\begin{table}[h]
 \begin{center}
  {\renewcommand{\arraystretch}{1.1}
  \begin{tabular}{| c | c  c  c  c  c  c  c  c  c  |}
     \hline
     \rule[5mm]{0mm}{0pt}
   & $Q$ & $u^{c}$ &
   $d^{c}$ & $L$ &
   $e^{c}$ & $\nu^{c}$ &
   $H_u$ & $H_d$ &
   $S$ \\
   \hline
   \rule[5mm]{0mm}{0pt}
$U(1)_A$& 2 & 2 & 4 &
   4 & 2 & 0 & $-4$ & $-6$ & 5 \\
   \rule[5mm]{0mm}{0pt}
$Z_{11}$& 2 & 2 & 4 &
   4 & 2 & 0 & 7 & 5 & 5 \\
\rule[5mm]{0mm}{0pt}
$Z_{12}$& 2 & 2 & 4 &
   4 & 2 & 0 & 8 & 6 & 5 \\
   \hline
  \end{tabular}
  }
  \caption{\footnotesize The anomalous $U(1)$ charge assignment for
  the DFSZ axion model.  Also shown are the charges under two discrete subgroups
  $Z_{11}$ and $Z_{12}$ which can stabilize the axion.}
  \label{charge5}
 \end{center}
\end{table}

\section{ Stabilizing the KSVZ  Axion}

The scalar sector of the non--supersymmetric KSVZ axion model \cite{KSVZ}
contains the SM doublet and a singlet field $S$.  All the SM
fermions are assumed to have zero PQ charge under the global
$U(1)_{PQ}$ symmetry.  The Yukawa sector
involving the SM fermions is thus unchanged.  An exotic quark--antiquark
pair $\Psi + \bar{\Psi}$ is introduced, which transform vectorially
under the SM (so the magnitude of
its mass term  can
be much larger than the electroweak scale), but has chiral transformations
under $U(1)_{PQ}$.  The QCD anomaly needed for the axion potential
arises from these exotic quarks.  The Lagrangian involving the
singlet field and these vector quarks is given by
\beq \Delta L=S\Psi\bar{\Psi}+h.c.
\eeq
When $S$ field develops a VEV, the PQ symmetry is spontaneously
broken leading to the axion in the light spectrum.

The global PQ $U(1)$ symmetry is susceptible to unknown quantum
gravity corrections.  We shall attempt to stabilize the KSVZ axion by
making use of discrete gauge symmetries with anomaly cancellation
by the Green--Schwarz mechanism.  The most dangerous non--renormalizable
term in the scalar potential that can destabilize the axion is
$S^n/M_{\rm Pl}^{n-4}$, as in the case of the DFSZ axion.  We seek
a discrete gauge symmetry that would forbid such terms.

In order for the Green--Schwarz mechanism for anomaly cancellation
to be viable, the anomaly coefficients $A_2$ and $A_3$ corresponding
to the $[SU(2)_L]^2 \times U(1)_A$ and $[SU(3)_C]^2 \times U(1)_A$
should equal each other at the $U(1)$ level.  This would imply that
the $\Psi + \bar{\Psi}$ fields can not all be singlets of $SU(2)_L$.
The simplest example we have found is the addition of a ${\bf 5}+
{\bf \bar{5}}$ of $SU(5)$ to the SM spectrum.  Such a modification
is clearly compatible with grand unification.  The ${\bf 5}$
contains a $({\bf3,1})$ and a $({\bf
1,2})$ under $SU(3)_C \times SU(2)_L$.  We allow the following Yukawa
coupling involving these fields:
\begin{equation}
{\cal L} \supset \lambda {\bf 5} {\bf \bar{5}} S + h.c.
\end{equation}
If we denote the PQ charges of ${\bf 5}$ and ${\bf \bar{5}}$ as
$\phi$ and $\bar{\phi}$, invariance of Eq. (16) under a surviving
discrete $Z_N$ symmetry would imply
\begin{equation}
\phi + \bar{\phi} + s = pN
\end{equation}
where $p$ is an integer.  In this simple model, all the SM particles
are assumed to be trivial under the PQ symmetry.
The discrete anomaly coefficients are then
$A_3=A_2=\frac{3}{2}(\phi+\bar{\phi})=\frac{3}{2}(p N-s)$.  Since
$A_2 = A_3$, the gauge anomalies are cancelled by the GS mechanism.
As long as $N \geq 10$, all dangerous couplings that would destabilize
the axion through non--renormalizable terms will be sufficiently
small.  We see that the KSVZ axion can be made consistent in a
simple way.

We have also examined the possibility of stabilizing the axion
by introducing only a single pair of fermions under the SM gauge group,
rather than under the grand unified group. Let us consider a class of models with
a pair of fermions transforming
under $SU(3)_{C}\times SU(2)_L\times U(1)_Y\times
U(1)_A$ as
\beq \Psi({\bf 3,n,}y,\psi)+\bar{\Psi}({\bf \bar{3},\bar{n}
},-y,\bar{\psi})~, \eeq along with a scalar field $S({\bf
1,1},0,s)$. The Lagrangian of this model contains a term $\Psi \bar{\Psi}S$
as in Eq. (15) and its invariance under an unbroken $Z_N$ symmetry imposes
the constraint
\beq \psi+\bar{\psi}+s=p N\eeq
where $p$ and $N$ are integers.
Since the SM particles all have zero anomalous $U(1)$ charge, the
anomlay coefficients arise solely from the
$(\Psi+\bar{\Psi})$ fields.  They are
\begin{eqnarray}
A_3&=&\frac{1}{2}(n\psi+n\bar{\psi})=\frac{n}{2}(pN-s)\nonumber\\
A_2&=&\frac{(n-1)n(n+1)}{12}(3\psi+3\bar{\psi})=\frac{(n-1)n(n+1)}{4}(pN-s).
\end{eqnarray}
The Green-Schwarz discrete anomaly
cancellation condition, Eq. (8), implies
\beq s=pN+\frac{2(-m+bm')}{n(b(n^2-1)-2)}\eeq
where  $b \equiv k_3/k_2$.

By choosing specific values of the Kac--Moody levels, one can solve for $s$,
the singlet charge. For
instance, in the simple case when $k_3=k_2\Leftrightarrow b=1$, \beq s=
\frac{2(m'-m)}{n^3-3n}N .
\eeq
We  have normalized all $U(1)_A$ charges to be integers, including $s$,
so the unbroken $Z_N$ symmetry will be transparent.

When $n=2$, $\Psi$ and $\bar{\Psi}$ are $SU(2)$ doublets. One can calculate
the charge of $S$ from Eq. (22) and determine the
allowed discrete symmetries.
For $b=1$, the solution is  $s=0~mod~ N$.  This solution
would imply that $S^n$ terms in the potential are allowed for any $n$,
in conflict with the axion solution.  A similar conclusion can be
arrived at for $b = 1/2$.  For other values of $b$, the $Z_N$
symmetry typically turns out to be too small to solve the
strong CP problem.  For example, if $b=(2,3,1/3,3/2)$, the allowed discrete
symmetries are $(Z_4, Z_7, Z_3, Z_5)$.  A special case occurs when
$b= 2/3$, in which case $s$ is undetermined, since $A_3/k_3 = A_2/k_2$.
If one chooses $s \geq 10$, the KSVZ axion can be stabilized in this case.

If the quarks $\Psi$ and $\bar{\Psi}$ are triplets of $SU(2)_L$, stability of the
DSVZ axion solution can be guaranteed in a simple way.
For $b\equiv k_3/k_2 = (1,2,3,1/2,1/3,2/3,3/2)$, which are the
allowed possibilities if we confine to Kac--Moody levels less than
3, we have from Eq. (21), the unbroken discrete symmetries to be
$(Z_9, Z_{21}, Z_{33}, Z_6, Z_3, Z_{15}, Z_{30})$ respectively.
For all $Z_N$ with $N \geq 10$, the axion solution will be stable
against quantum gravitational corrections.

\section{SUSY extensions of invisible axions and a natural link
between $M_{\rm SUSY}, f_a$ and the $\mu$ term}

The models of the previous sections can be easily generalized to
incorporate supersymmetry.  The DFSZ axion is a natural extension
of SUSY, since supersymmetry requires the existence of two Higgs
doublets.  For the axion to be weakly coupled (or invisible), we
also need a pair of singlet scalars $S$ and $\tilde{S}$.  As we
shall see, there is a natural link between the SUSY breaking scale
$M_{\rm SUSY}$, the PQ symmetry breaking scale $f_a$, and the
supersymmetric $\mu$--term.

\subsection{SUSY DSFZ axion}

The superpotential of the model is taken to be
\beq \label{sp2}
W=Qu^{c}H_u+Qd^{c}H_d+Le^{c}H_d+L\nu^{c}H_u+M_R\nu^{c}\nu^{c}+\lambda_1\frac{H_u
H_d S^2}{M_{\rm Pl}} +\lambda_2 \frac{(S\tilde{S})^{2}}{M_{\rm Pl}}.
 \eeq
As discussed earlier, the VEVs of the singlet fields $S$ and
$\tilde{S}$ set $f_a \sim 10^{11}$ GeV.  Consequently the
$\lambda_1$ term in Eq. (23) will induce a $\mu$--term for the
MSSM Higgsinos given by \cite{mu}
\beq \label{gmu2} \mu=\lambda_1\frac{\left\langle S\right\rangle^2}
{M_{\rm Pl}}\approx 10^2~
\mathrm{GeV}. \eeq
Furthermore, we have the following relation for the SUSY breaking mass terms
in terms of the axion decay constant:
\begin{equation}
M_{\rm SUSY} \simeq {f_a^2 \over M_{\rm Pl}}.
\end{equation}
Thus, three a priori unrelated quantities get linked in these models.

To see the link between $M_{\rm SUSY}$ and $f_a$, we need to minimize
the scalar potential\footnote{As far as the VEVs of the $S$ and $\tilde{S}$
are concerned, we can ignore the $H_{u}H_{d}S^2/M_{\rm Pl}$ term, which can
affect the VEVs by only
$10^{-14}$ GeV.} which contains both the soft SUSY breaking terms and
the F-terms:
\beq V=
(\lambda_{2}C\frac{(S\tilde{S})^2}{M_{\rm Pl}}+h.c)+{m_s}^2|S|^2+{m_
{\tilde{S}}}^2{|\tilde{S}|}^2+4\lambda_2\frac{|S\tilde{S}|^2}{M_{\rm Pl}^2}
(|S|^2+{|\tilde{S}|}^2).\eeq
Here $C$ is the soft SUSY breaking
term corresponding to the non--renormslizable coupling in the superpotential of
Eq. (23).\footnote{Notice that C can be either positive or negative.  We shall follow
a notation where C is positive. By
setting $S=|S|e^{i\alpha}$ and $\tilde{S}=|\tilde{S}|e^{i\beta}$,
$C$ would appear in the potential as
$2\lambda_{2}C\frac{|S\tilde{S}|^2}{M_{Pl}}\cos{2(\alpha+\beta)}$.
Minimizing this potential we obtain
$\cos{2(\alpha+\beta)}=-1$.}
Let us assume for simplicity that $m_S = m_{\tilde{S}}$.  Relaxing this
assumption does not have any major effects.

By minimizing the potential of Eq. (26)  \cite{babumohap} we obtain
\begin{equation}\label{vev}
x^2=\frac{C\pm\sqrt{C^2-12{m_{S}}^2}}{12\lambda_2}M_{\rm Pl}\\
\end{equation}
where $x^2=\left\langle |S|\right \rangle^2=\left\langle
|\tilde{S}|\right\rangle^2$.
This is the desired relation between $f_a$ and$M_{\rm SUSY}$ since Eq. (27)
implies
\beq f_a \sim
\left\langle |S| \right\rangle \sim \sqrt{M_{\rm Pl}M_{\rm SUSY}}
 \sim 10^{11}~GeV .\eeq

The scalar spectrum of this model contains two  Higgs bosons
from $S$ and $\tilde{S}$, besides those in the MSSM, with masses given by
\begin{equation}
{m_{S_1}}^2= \frac{-2{\lambda_2}x^2 C{M_{\rm Pl}}+24 {\lambda_2}^2x^4}
{{M_{\rm Pl}}^2},~~~
{m_{S'_2}}^2= \frac{6x^2{\lambda_2}x^2 C{M_{\rm Pl}}+40{\lambda_2}^2x^4}
{{M_{\rm Pl}}^2}~
\end{equation}
where $x$ is given in Eq. (27).  Note that both scalars have masses
of order $M_{\rm SUSY}$.  Their mixings with the doublet Higgs is
very small, being suppressed by a factor $v/x \sim 10^{-8}$.

Among the pseudoscalars in $S, \tilde{S}$, one combination
remains massless and is identified as the axion.  (QCD anomaly induces a tiny
mass for the axion.)  The orthogonal combination has a mass of order
$M_{\rm SUSY}$, given by
\beq
m_{S_3}^2 = \frac{4{\lambda_2}C{x}^2}{M_{\rm Pl}} \eeq

The  fermionic partner of the axion, the axino ($\tilde{a}$),
receives a mass after SUSY breaking from two sources \cite{axino}.
The superpotential
terms induce an axino mass of order $v_uv_d/M_{\rm Pl} \sim 10^{-5}$ eV.
A second source of its mass is through a Lagrangian term
 \beq  {\cal L} \supset \int d^{4}\theta
(S^{\dagger}S)^2/{M_{Pl}}^{2} \longrightarrow \tilde{a} \tilde{a}
F^{*}_{S}\left\langle S^{*}\right\rangle/{M_{\rm Pl}}^{2}\sim
m_{\rm SUSY}^2/M_{\rm Pl} \sim 10^{-3}~ eV. \eeq
The fermionic component of the heavy pseudoscalar $S_3$
obtains a mass from the superptential, of order $M_{\rm SUSY}$:
\beq \int d^{2}\theta
(S\tilde{S})^2/{M_{\rm Pl}} \longrightarrow
S_{F}S_{F}\left\langle \tilde{S}\tilde{S}\right\rangle/{M_{\rm Pl}}
\sim M_{\rm SUSY} S_F S_F~. \eeq

The axino is then the lightest SUSY particle in this model.  The lightest
neutralino (the LSP in conventional SUSY models)
will decay eventually into an axino.  To estimate this decay lifetime,
we first note that the doublet Higgsino--axino mixing angle $\theta$
is of order $M_{\rm SUSY}/M_{\rm Pl})^{1/2} \sim 10^{-8}$.  Suppose that
the lightest neutralino is mostly a Bino and that its mass is larger
than the lightest Higgs boson mass.  The decay $\tilde{B} \rightarrow
\tilde{a} + h^0$ can then proceed with a rate estimated to be
$\Gamma \sim (3\alpha_1/20)(\theta/\tan\beta)^2 M_{\tilde{B}}$.  For
reasonable values of the parameters we see that the lifetime is of
order $10^{-7}$ seconds, so the decay will  likely occur outside the detector.
For some other  values of the model parameters, the decay length could be
of order cm, which should be measurable experimentally.

Turning now to the stability of the SUSY DFSZ axion, as in the non--SUSY
case we make use of a discrete $Z_N$ symmetry.  We start with the same
$U(1)_A$ charge assignment as in Table 2.
One difference in the SUSY extension\footnote{The
Lagrangian actually has four $U(1)$ symmetries, hypercharge,
Baryon number, Peccei-Quinn and $R$ symmetries. However, the $U(1)_R$
is broken by the gaugino mass term to a discrete subgroup.
Conventionally, the gauginos are assigned $R$ charge of 1,
after the $F$--component of a spurion field $Z$ acquires
a VEV, to generate gaugino masses, the $R$ symmetry is reduced to a
$Z_2$.  This $Z_2$ will be related to a subgroup of
$B-L$ symmetry \cite{wk}.} is that there is an extra
$\tilde{S}$ singlet field.  The charge assignment
under a $Z_{N}$ subgroup is displayed in Table 3 for $N=22$.
The $Z_{11}$ symmetry of Table 2 has been embedded into a bigger
group $Z_{22}$ in order to prevent a bare mass term $S\tilde{S}$ in
the superpotential, while allowing a term $(S \tilde{S})^2$.
This enables us to relate the $\mu$ term, $M_{\rm SUSY}$
and $f_a$.  This $Z_{22}$ symmetry forbids to high orders
all dangerous terms of the type
\beq \label{ax22} \frac{S^m \tilde{S}^{n-m}}{M_{Pl}^{n-3}}.
 \eeq
that can potentially destabilize the axion.

A second point that distinguishes the SUSY DFSZ axion model
from non-SUSY case is
that the anomaly coefficients now receive contributions from
the Higgsino and the gaugino.  (The absence of $S\tilde{S}$ term in
$W$, with $(S\tilde{S})^2$ allowed implies that the Grassman variable
$\theta$ must transform non-trivially under the $Z_N$ symmetry.)
Note that the anomalies of $Z_{22}$ are cancelled by Green--Schwarz mechanism.

In Table 3  we have also displayed a $Z_2$ symmetry
that turns out to be an unbroken subgroup of $B-L$, which is
clearly anomaly free.  This $Z_2$ symmetry serves as $R$--parity
needed for the stability of the proton in MSSM \cite{wk}.

\begin{table}[h]
 \begin{center}
  {\renewcommand{\arraystretch}{1.1}
  \begin{tabular}{| c | c  c  c  c  c  c  c  c  c c c |c |}
     \hline
    \rule[5mm]{0mm}{0pt}
   & $Q$ & $u^{c}$ &$d^{c}$ & $L$ & $e^{c}$ & $\nu^{c}$ &   $H_u$ & $H_d$ &   $S$& $\tilde{S}$&$\alpha$&$(A_2,~A_3)$\\
   \hline
    \rule[5mm]{0mm}{0pt}
$Z_{22}$& 3 & 19 & 1 &11&
   15 & 11& 22 & 18 & 13 & 20 & 11&(19,~8)\\
   \rule[5mm]{0mm}{0pt}
    $Z_{2}$& 1 & 1 & 1 &1&
   1 & 1& 2 & 2 & 2 & 2 & 2&(0,~0)\\

   \hline
  \end{tabular}
  }
  \caption{\footnotesize The charges of  particle corresponding SUSY DSFZ model
  with a $Z_{22}$
  discrete gauge symmetry. }
  \label{charge}
 \end{center}
\end{table}

\subsection{SUSY KSVZ axion model and its stability}

In the SUSY version of the KSVZ axion model we assume the
superpotnetial to be
\beq
W=Qu^{c}H_u+Qd^{c}H_d+Le^{c}H_d+L\nu^{c}H_u+M_R\nu^{c}\nu^{c}+\Psi\bar{\Psi}S+\frac{(S\tilde{S})^2}
{M_{Pl}}.
 \eeq
Here $\Psi + \bar{\Psi}$ are the new vectorlike fields with QCD anomaly.
We use the same mechanism as in the DFSZ model to relate $f_a$ with
$M_{\rm SUSY}$ and $\mu$ terms.  It requires that the
mass term $S\tilde{S}$ be absent in $W$, but the coupling $(S \tilde{S})^2$
be present.

The simplest possibility is to assume that the feilds $\Psi + \bar{\Psi}$
belong to ${\bf 5}+{\bf \bar{5}}$ of $SU(5)$.  Gauge coupling unification
that seems to occur within MSSM can be maintained with the introduction
of complete multiplets of $SU(5)$.  The charge assignment under a $Z_{12}$
discrete symmetry are shown in Table 4.  As before, the anomalies
of $Z_{12}$ vanish via the discrete GS mechanism. The lowest order term
that breaks the PQ symmetry is $S^{12}$, which is harmless for the axion
solution.

\begin{table}[h]
 \begin{center}
  {\renewcommand{\arraystretch}{1.1}
  \begin{tabular}{| c | c  c  c  c  c  c  c  c  c c c c |c|}
     \hline
     \rule[5mm]{0mm}{0pt}
   & $Q$ & $u^{c}$ &
   $d^{c}$ & $L$ &
   $e^{c}$ & $\nu^{c}$ &
   $H_u$ & $H_d$ &$\Psi+\bar{\Psi}$&
   $S$& $\tilde{S}$&$\alpha$& $(A_2,~A_3)$\\
   \hline
   \rule[5mm]{0mm}{0pt}
$Z_{12}$& 0 & 0 & 0 &0&
   0 & 0& 0 & 0 & 11 & 1 & 5&6&(5,~11)\\
   \rule[5mm]{0mm}{0pt}
    $Z_{2}$& 1 & 1 & 1 &1&
   1 & 1& 2 & 2 & 2 & 2 & 2&2&(0,~0)\\
   \hline
  \end{tabular}
  }
  \caption{\footnotesize SUSY KSVZ axion model realized from $Z_{12}\times Z_2$
  symmetry.}
 \end{center}
\end{table}

In Table 5 we have displayed a variant KSVZ axion model where the
Green--Schwarz anomaly cancellation occurs at a higher Kac--Moody
level with $k_3 = 3,$ and $k_2 = 2$.  We have extended the
solution obtained for non--SUSY case to the SUSY case here.
There is a single $\Psi+\bar{\Psi}$ field, belonging to
a  $({\bf 3},{\bf 3})$ representation.  Exact $R$--parity
is also realized as a $Z_2$ symmetry, as shown in Table 5.

\begin{table}[h]
 \begin{center}
  {\renewcommand{\arraystretch}{1.1}
  \begin{tabular}{| c | c  c  c  c  c  c  c  c  c c c c |c|}
     \hline
     \rule[5mm]{0mm}{0pt}
   & $Q$ & $u^{c}$ &
   $d^{c}$ & $L$ &
   $e^{c}$ & $\nu^{c}$ &
   $H_u$ & $H_d$ &$\Psi+\bar{\Psi}$&
   $S$& $\tilde{S}$&$\alpha$& $(A_2,~A_3)$\\
   \hline
   \rule[5mm]{0mm}{0pt}
$Z_{30}$& 0 & 0 & 0 &0&
   0 & 0& 0 & 0 & 29 & 1 & 14&15&(24,~57/2)\\
   \rule[5mm]{0mm}{0pt}
    $Z_{2}$& 1 & 1 & 1 &1&
   1 & 1& 2 & 2 & 2 & 2 & 2&2&(0,~0)\\
   \hline
  \end{tabular}
  }
  \caption{\footnotesize SUSY KSVZ axion realized at higher Kac--Moody level
  from $Z_{30}\times Z_2$ symmetry. }
 \end{center}
\end{table}

In conclusion, we have shown how the axion solution of the strong
CP problem can be stabilized against potential quantum gravitational
corrections via discrete gauge symmetries.  Green--Schwarz anomaly
cancellation mechanism plays a crucial role in our models.  Both the
DFSZ axion model and the KSVZ axion model are realized in rather
simple ways without enlarging the low energy particle content.
In the supersymmetric extension of these models, we have found
an interesting link between the SUSY breaking scale, the axion
decay constant, and the SUSY--preserving $\mu$--term.

\section{Acknowledgement}

We thank Ts. Enkhbat and J. Lykken for useful discussion. This
work is supported in part by DOE Grant \# DE-FG03-98ER-41076, a
grant from the Research Corporation and by DOE Grant \#
DE-FG02-01ER-45684.

\end{document}